\documentclass[
showpacs,
 amsmath,amssymb,
 aps,superscriptaddress,
 10pt,
twocolumn
]{revtex4-1}

\usepackage[usenames,dvipsnames]{color}
\usepackage{graphicx,textcase} 
\usepackage{dcolumn,overpic} 
\usepackage{bm,color}
\usepackage[T1]{fontenc}
\usepackage{ae,aecompl}
\usepackage{chngcntr,gensymb}
\usepackage[para]{threeparttable}
\usepackage{etoolbox}
\usepackage{lipsum}
\usepackage[bottom]{footmisc}
\usepackage{setspace}

\AtEndEnvironment{table*}{\vskip10pt}{}{}

\usepackage{url}
\usepackage{hyperref}
\hypersetup{colorlinks=true,breaklinks,linkcolor=blue,urlcolor=blue,citecolor=blue}

\begin{document}

\title{Relativistic exchange interactions in CrX$_3$ (X=Cl, Br, I) monolayers}
\author{Y.~O.~Kvashnin}
\affiliation{Uppsala University, Department of Physics and Astronomy, Division of Materials Theory, Box 516, SE-751 20 Uppsala, Sweden}
\author{A. Bergman}
\affiliation{Uppsala University, Department of Physics and Astronomy, Division of Materials Theory, Box 516, SE-751 20 Uppsala, Sweden}
\author{A.~I. Lichtenstein}
\affiliation{Institute of Theoretical Physics, University of Hamburg, Jungiusstrasse 9, 20355 Hamburg, Germany}
\affiliation{European X-Ray Free-Electron Laser Facility, Holzkoppel 4, 22869 Schenefeld, Germany}
\affiliation{The Hamburg Centre for Ultrafast Imaging, Luruper Chaussee 149, 22761 Hamburg, Germany}
\author{M.~I. Katsnelson}
\affiliation{Radboud University of Nijmegen, Institute for Molecules and Materials, Heijendaalseweg 135, 6525 AJ Nijmegen, The Netherlands}

\date{\today}

\begin{abstract}
It has been predicted theoretically and indirectly confirmed experimentally that single-layer CrX$_3$ (X=Cl, Br, I) might be the prototypes of topological magnetic insulators (TMI). 
In this work, by using first-principles calculations combined with atomistic spin dynamics we provide a complete picture of the magnetic interactions and magnetic excitations in CrX$_3$.
The focus is here on the two most important aspects for the actual realization of TMI, namely the relativistic magnetic interactions and the finite-size (edge) effects.
We compute the full interaction tensor, which includes both Kitaev and Dzyaloshinskii-Moriya terms, which are considered as the most likely mechanisms for stabilizing topological magnons.
First, we instigate the properties of bulk CrI$_3$ and compare the simulated magnon spectrum with the experimental data [Phys. Rev. X 8, 041028 (2018)].
Our results suggest that a large size of topological gap, seen in experiment ($\approx$ 4 meV), can not be explained by considering pair-wise spin interactions only.
We identify several possible reasons for this disagreement and suggest that a pronounced magneto-elastic coupling should be expected in this class of materials.
The magnetic interactions in the monolayers of CrX$_3$ are also investigated.
The strength of the anisotropic interactions is shown to scale with the position of halide atom in the Periodic Table, the heavier the element the larger is the anisotropy. 
Comparing the magnons for the bulk and single-layer CrI$_3$, we find that the size of the topological gap becomes smaller in the latter case.
Finally, we investigate finite-size effects in monolayers and demonstrate that the anisotropic couplings between Cr atoms close to the edges are much stronger than those in ideal periodic structure.
This should have impact on the dynamics of the magnon edge modes in this class of materials.
\end{abstract}

\maketitle

\section{Introduction}

Magnetism in two dimensions (2D) has essential peculiarities in comparison with the conventional 3D case and in general should be strongly suppressed. 
The celebrated Mermin-Wagner theorem\cite{PhysRevLett.17.1133} tells us that the long-range magnetic order seizes to exist in 2D at any finite temperature being completely destroyed by thermal spin fluctuations.
However, this result was obtained theoretically assuming isotropic short-range magnetic interactions between the atomic spins (Heisenberg model).
At the same time, the famous Onsager solution to the 2D Ising model shows ordering at finite temperature\cite{PhysRev.65.117}. In real materials, the out-of-plane single-site or two-site anisotropy will result in the same behavior, due to opening the gap in the spin wave spectrum and the related suppress of thermal fluctuations; however, the ordering temperature is much lower than the strength of exchange interactions in this case\cite{PhysRevB.60.1082}.

There have been reports on observed magnetism in graphene, which is the most famous 2D material.
However, it usually originates from defects, e.g. vacancies\cite{PhysRevLett.117.166801}, edges\cite{RevModPhys.88.025005, Slota2018} or adatoms\cite{Gonzalez-Herrero437, Nair2012}.
Thus, a rather high density of defects is required to result in a magnetic ordering\cite{Cervenka2009}.
At the same time, the magnetism in layered van-der-Waals (vdW) bonded systems has been known since long time ago\cite{doi:10.1143/JPSJ.15.1664, doi:10.1063/1.1714194}.
However, the first reports of successful exfoliation of these materials down to monolayer thickness appeared only in 2017\cite{Gong2017, Huang2017}. 
Chromium triiodide (CrI$_3$) was among the first materials, where magnetism was observed even within a single layer.
This has sparked an enormous interest in these materials and many more 2D magnets have been synthesized from the bulk vdW-bonded structures\cite{Burch2018, Gibertini2019}.
Even room-temperature ferromagnetism has been reported\cite{Bonilla2018}, however it has recently been reported to be suppressed by charge density wave formation\cite{fumega2019absence}.

In chromium trihalides, the magnetic ions (Cr$^{3+}$) within each layer form a honeycomb lattice\cite{cri3-struct, cryst7050121}.
In this lattice, topological magnons have been proposed to exist theoretically, due to the appearance of conical point in magnon spectrum in isotropic exchange approximation\cite{Owerre_2016, PhysRevB.97.081106}. The relativistic interactions can prevent the conical magnon band crossing and from the gap between two branches. 
Recently, an inelastic neutron scattering experiment was performed on bulk CrI$_3$, which showed a pronounced gap of about 4 meV opening between the two magnon branches\cite{PhysRevX.8.041028}. 
The latter implies that the two modes get a topological character and hence this material might be one of the first realizations of the \textit{topological magnon insulators} (TMIs)\cite{PhysRevB.87.144101}.
If this is the case, then it means that one can excite topologically protected edge spin waves, which will be of enormous importance for magnonic applications.
Chen \textit{et al.}\cite{PhysRevX.8.041028} have attributed the opening of topological gap to the Dzyaloshinskii-Moriya (DM) interaction\cite{DZYALOSHINSKY1958241, PhysRev.120.91}, which is the most discussed source for the topological magnons\cite{PhysRevB.87.144101, Owerre_2016, PhysRevB.97.081106, PhysRevB.90.024412}.
However, the peculiarity in CrI$_3$ is that the nearest-neighbour (NN) DM vector is forbidden by symmetry and only for the next NN one it is finite.
In order to fit their data, the authors of that work had to use the value of the DM vector, which was larger than that of the corresponding Heisenberg interaction. 
At the same time, it was shown in Ref.~\cite{PhysRevLett.124.017201} that the same effect can be achieved by considering the Kitaev interaction\cite{KITAEV20062}, which is common for systems with edge-sharing octahedra\cite{PhysRevLett.102.017205}.
This term, contrary to the DM interaction, is allowed by symmetry between nearest neighbours of the honeycomb lattice.
Again, extremely large values of this term are necessary to reproduce the data of Chen \textit{et al}\cite{PhysRevX.8.041028}.
Both DM and Kitaev interactions originate from spin-orbit coupling (SOC) and therefore they are usually small in $3d$-based materials (smaller than the Heisenberg exchange). However, in the case of CrI$_3$ it is the orbitals of iodine that might be the source of anisotropic interactions, and SOC on heavy iodine ion may be not so small.
Still, as one can see, the true origin of such a large gap in the spin-wave dispersion is not certain.

First principles calculations, primarily based on density functional theory (DFT) play a crucial role in the field of 2D magnets. 
There has been a number of predictions for new promising materials and some of them were later realized experimentally (see e.g. Refs.~\onlinecite{PhysRevB.79.115409, PhysRevX.3.031002, PhysRevB.88.201402, Ma-ACSNano, cri3-struct}). 
Bulk chromium trihalides are insulators with the Cr magnetic moment value being close to the nominal $S=\frac{3}{2}$ or 3 $\mu_B$\cite{doi:10.1063/1.1714194}. 
This implies that the states associated with magnetism are well localised and the system can be well described by an effective Heisenberg model of interacting spins. 
This kind of studies have been performed focusing on the effective exchange interactions and magnetic anisotropy, which were computed from the DFT total energies\cite{PhysRevB.100.205409, Lado_2017, PhysRevB.98.144411, olsen_2019}.

The isotropic pair-wise exchange interactions in bulk CrCl$_3$ and CrI$_3$ were computed from first principles and analysed in detail by Besbes \textit{et al.}\cite{PhysRevB.99.104432}.
The authors employ two different electronic structure codes and analyse the difference in the exchange interactions, computed using magnetic force theorem (MFT)\cite{jij-1}.
The authors demonstrate that there are competing ferromagnetic and antiferromagnetic contributions to the NN exchange coupling, which is related to a complex interplay between local exchange and crystal field splitting.
This is related with the fact that the Cr-X-Cr bond angle is close to 90$^{\circ}$, which makes it hard to predict its sign even from the Goodenough-Kanamori rules\cite{KANAMORI195987}.
Furthermore they showed that the MFT applied on the Cr states only leads to the antiferromagnetic (AFM) NN coupling in CrCl$_3$, which contradicts the total energy results.
They argue that the magnetism of ligand states plays a crucial role in stabilizing the ferromagnetic state in these systems.

Kashin \textit{et al}\cite{Kashin_2020} have analysed the magnetic interactions in monolayered CrI$_3$, using similar methods as in Ref.~\onlinecite{PhysRevB.99.104432}.
They performed orbital decomposition of the exchange couplings and studied their dependence on the strength of local Coulomb interaction $U$.
The authors estimated the magnetic ordering temperature and found it to be in good agreement with experiment. 
Finally, they simulated the magnon spectrum of monolayered CrI$_3$ and showed how it can be altered by applying bias voltage.

Refs.~\onlinecite{PhysRevB.99.104432, Kashin_2020} neglect the effect of spin-orbit coupling, which only allowed them to address isotropic (Heisenberg) exchange interactions.
In this work we do the natural next step and provide a complete first-principles description of the relativistic magnetic interactions in CrX$_3$ systems.

The paper is organized as follows. First, we present in Section II the details of the performed calculations and describe the formalism for computing relativistic magnetic interactions. 
Section III contains the main results and is divided in several sub-sections.
First, we present the magnetic properties of bulk CrI$_3$, show simulations of the spin wave dispersion and compare it with the experimental data of Chen \textit{et al}\cite{PhysRevX.8.041028}. 
After that, we do the same analysis for the series of CrX$_3$ monolayers (X=$\{$Cl,Br,I$\}$) and discuss how the change of ligand ion affects the relativistic exchange interactions.
Finally, we discuss how finite-size effects can drastically affect the magnetic couplings in CrI$_3$. 
In section IV we discuss our main findings and outline potential directions for the future work.

\section{Computational details}

\subsection{Electronic structure}

Electronic structure was described by means of DFT and its extensions to account for strong local Coulomb repulsion $U$.
The $U$ term was added in two different ways: either on top of non-spin-polarized\cite{ldau1} or spin-polarized\cite{lsdau} exchange-correlation (xc) functional.
To distinguish the two, we refer to them as DFT+$U$ and sDFT+$U$ (where $s$ stands for spin-polarized version of the xc functional, see e.g. Refs.~\onlinecite{PhysRevB.91.241111, millis-chen-2016}).
In the former case the magnetism is added specifically on the manifold of states, where $U$ correction is applied (Cr-$3d$ orbitals in our case).
Thus, there is no intrinsic magnetism of the ligand (halide) states, which, on the contrary, appears in sDFT(+$U$) case.
DFT+$U$ and sDFT+$U$ have been previously shown to produce substantially different results for the total energies\cite{millis-chen-2016} and the exchange interaction parameters\cite{PhysRevB.97.184404}.
As was mentioned above, the magnetism of ligands is particularly important in CrX$_3$ systems, which motivated us to compare the results obtained with these two flavours of $+U$ methods.

First, we considered bulk CrI$_3$, for which we used experimental structural parameters taken from Ref.~\onlinecite{cri3-struct}.
Next, we constructed a series of monolayered CrX$_3$ compounds. 
Due to 3D periodic boundary conditions, we have added a vacuum of about 20$\AA$ between the layers to avoid spurious interactions between them.
The crystal structures were then relaxed in VASP\cite{paw,vasp}.
For this purpose we used PBESol\cite{PhysRevLett.100.136406} xc functional, which often gives better structural properties as compared to the original PBE\cite{gga-pbe}.
The plane-wave energy cut-off was set to 350 eV along with 17$\times$17$\times$1 k-point grid.
The forces on each atom were minimized down to 0.001 eV/\AA.
Once the structures were optimized, calculations of the magnetic interactions were calculated with PBE using full-potential linear muffin-tin orbital-based code RSPt\cite{rspt-web,rspt-book}.
More details about these calculations will be given in the next section.

\subsection{Magnetic interactions}

In order to describe the magnetic interactions in CrX$_3$ systems, we considered a generalized Heisenberg Hamiltonian:
\begin{eqnarray}
\hat H = -\sum_{i \ne j} \sum_{\{\alpha,\beta\}=\{x,y,z\}} e^{\alpha}_i J^{\alpha \beta}_{ij} e^{\beta}_j ,
\label{heis}
\end{eqnarray}
where $e^{\alpha}_i$ is the $\alpha$ component of the unitary vector pointing along the direction of the spin located at the site $i$.
Inter-site magnetic interactions are rank-2 tensors and are denoted as $J^{\alpha \beta}_{ij}$.
The antisymmetric part of this tensor can be rewritten in terms of DM vector, having e.g. such $z$-component:
\begin{eqnarray}
\vec D^z_{ij} = (J^{xy}_{ij} - J^{yx}_{ij})/2.
\label{dz}
\end{eqnarray}

In addition, we also considered local magnetic anisotropy term:
\begin{eqnarray}
\hat H_{\mathrm{anis}} = - \sum_i K_{zz} e^2_{i,z},
\end{eqnarray}
where $K_{zz}$ is uniaxial magnetic anisotropy constant, which if positive sets the magnetization direction out of plane (perpendicular to the monolayer).
We have extracted this term from the magnetocrystalline anisotropy energy (MAE), which is defined as the total energy difference between in-plane and out-of-plane directions of the magnetization.
Note that the $K_{zz}$ and MAE are not in one to one correspondence in our model, since the latter is also influenced by anisotropic exchange interactions (such as the difference between $J^{xx}$ and $J^{zz}$).

The parameters in Eq.~\eqref{heis} for a given real material can be extracted by means of MFT.
This approach is essentially a linear-response theory formulated for second order perturbation in the deviations of spins from equilibrium magnetic configuration.
It is based on a mapping procedure of the tilting of classical spins, coupled via Eq.~\eqref{heis} and the electronic Hamiltonian.
The MFT was originally formulated for the case of isotropic Heisenberg interactions in the absence of spin-orbit coupling\cite{jij-1,PhysRevB.61.8906}, but later it was extended to take into account relativistic effects to allow to compute the full interaction tensor\cite{ANTROPOV1997336,PhysRevB.68.104436, PhysRevB.79.045209, SECCHI201561}.
We present a derivation of the formulae based on Green's functions formalism below.

We begin by perturbing the spin system by deviating the initial moments ($\vec e_0$) on a small angle $\vec{\delta \varphi}$ (the site index is omitted at the moment):
\begin{eqnarray}
\vec e = \vec e_0 + \delta \vec e + \delta^2 \vec {e} = \vec e_0 + \bigl[ \vec{\delta \varphi} \times \vec e_0 \bigl] -\frac{1}{2} \vec e_0 (\vec {\delta \varphi})^2
\end{eqnarray}
Then one can write the Hamiltonian (Eq.~\eqref{heis}) of the perturbed system in terms of series in the order of $\vec \delta \varphi$:
\begin{eqnarray}
\hat H' = \hat H^{(0)} + \hat H^{(1)} + \hat H^{(2)}.
\end{eqnarray}

In the collinear limit, all spins point along the same direction, which we set parallel to $Z$ axis.
Then the tilting vectors have the following components:
\begin{eqnarray}
\vec{\delta\varphi} = (\delta\varphi^x ; \delta\varphi^y ; 0)  \\
\bigl[ \vec{\delta\varphi} \times \vec e_0 \bigl] = (\delta \varphi^y ; - \delta\varphi^x ; 0)
\end{eqnarray}

Focusing on the energy contributions of the second order in $\vec{\delta \varphi}$ (i.e. $\hat H^{(2)}$), we obtain:
\begin{widetext}
\begin{equation}
H^{(2)} = -\sum_{i \ne j} \biggl( J^{xx}_{ij} \delta\varphi_i^y \delta\varphi_j^y + J^{yy}_{ij} \delta\varphi_i^x \delta\varphi_j^x - J^{xy}_{ij} \delta\varphi_i^y \delta\varphi_j^x - J^{yx}_{ij} \delta\varphi_i^x \delta\varphi_j^y -\frac12 J^{zz}_{ij} ((\vec{\delta \varphi_i})^2 + (\vec{\delta \varphi_j})^2) \biggl)
\end{equation}
\end{widetext}

Then one can do the same perturbation for the electronic Hamiltonian ($\mathcal{H}$), which will become:
\begin{eqnarray}
\hat{\mathcal{H}'} = \hat U^\dagger  \hat{\mathcal{H}} \hat U = \hat{\mathcal{H}}^{(0)} + \hat{\mathcal{H}}^{(1)} + \hat{\mathcal{H}}^{(2)},
\end{eqnarray}
where $\hat U=\exp{(i\vec{\delta\varphi} \hat{\vec{\sigma}}}/2)$ and $\hat{\vec{\sigma}}$ is the vector of Pauli matrices.
The corresponding terms proportional to $\vec{\delta\varphi}$ can be identified and mapped onto generalized Heisenberg model. 
This way, the expressions for various components of $J^{\alpha\beta}_{ij}$ are obtained (now the large square brackets refer to the commutator):
\begin{widetext}
\begin{eqnarray}
J^{xx}_{ij}= \frac{\text{T}}{4}\sum_n \text{Tr}_{L,m} \bigl[ \hat{\mathcal{H}}_i , \hat \sigma^{y} \bigl] G_{ij}(i\omega_n)   \bigl[ \hat{\mathcal{H}}_j , \hat \sigma^{y} \bigl] G_{ji} (i\omega_n) \\
J^{xy}_{ij}= -\frac{\text{T}}{4}\sum_n \text{Tr}_{L,m}  \bigl[ \hat{\mathcal{H}}_i , \hat \sigma^{y} \bigl] G_{ij}(i\omega_n)   \bigl[ \hat{\mathcal{H}}_j , \hat \sigma^{x} \bigl] G_{ji} (i\omega_n)
\label{Jfrel2}
\end{eqnarray}
\end{widetext}
and similar expressions for $J_{ij}^{yy}$ and $J_{ij}^{yx}$. 
In the above equations, the summation is done over the Matsubara frequencies ($\omega_n$) and the trace is over the orbital indices denoted by $m$.
Note that $J^{xz}_{ij}$, $J^{zx}_{ij}$, $J^{yz}_{ij}$, $J^{zy}_{ij}$ are not of the second order in the tilting angles. 
Thus, for M||Z, only $D_z$ component (Eq.~\eqref{dz}) can be computed, while $D_x$ and $D_y$ are extracted from two additional calculations with the magnetization pointing along $X$ and $Y$, respectively.
This has been discussed in Ref.~\onlinecite{PhysRevB.68.104436}.

The present formalism has been implemented into RSPt, extending its previous capabilities to compute exchange interactions in correlated systems~\onlinecite{jij-2}. 
In case of (s)DFT+$U$, $\mathcal{H}_i$ in the above equations contain the DFT Hamiltonian as well as the static self-energy from the $U$ term.
Note that the present approach can also be used to study magnetic interactions in the presence of dynamical local correlations following Refs.~\onlinecite{PhysRevB.61.8906,SECCHI201561}.

In order to be able to compute the interaction between two sites, it is essential to define the orthogonal set of localised orbital on each of them. 
As was pointed out in Ref.~\onlinecite{PhysRevB.99.104432}, it is essential for CrX$_3$ systems to capture the delocalization of the Cr-$3d$ states in order to obtain proper description of the magnetic interactions. 
There is no unique recipe to the choose the projection and here we have adopted L{\"o}wdin-orthogonalized orbitals, which can be easily constructed in RSPt.
In principle, this orthogonalization should produce the states, which are relatively delocalized in real-space, making them more similar to Wannier orbitals.

Once the $J^{\alpha\beta}_{ij}$ tensor were extracted, they were used to compute adiabatic magnon spectra. The latter was done using UppASD software\cite{uppasd,uppasd-book}.

\section{Results and Discussion}

\subsection{CrI$_3$ bulk}

We computed the full tensor of exchange interactions, using the formalism described in Section IIB. 
In the most general case there are nine independent components of these interactions with each neighbour.
There are certainly bonds which are related by symmetry, but then the interaction tensor has to be transformed using the symmetry operation bringing one bond to another.
Therefore, plotting the components of the interaction with each particular neighbour is not that informative.
In order to facilitate the analysis and present the results in the most compact way, we have defined the following quantities:
\begin{eqnarray}
\bar J = (J^{xx}+J^{yy}+J^{zz})/3, \\
\mid \vec D \mid = \sqrt{(D^x)^2 + (D^y)^2 + (D^z)^2}, \\
J^S = \frac12 (J^{\alpha\beta}+J^{\beta\alpha})- \bar J,
\label{def-int}
\end{eqnarray}
where $\bar J$ represents the isotropic (Heisenberg) part of the interaction, $\mid \vec D \mid $ is the magnitude of the DM vector. 
The $J^S$ is the symmetric anisotropic part of the exchange coupling. 
In principle, $J^S$ contains both two-site anisotropy and the Kitaev term. In a certain reference frame, the former appear as off-diagonal elements and the latter as purely diagonal one.
In order to analyse all bonds at ones, we show the three eigenvalues of the $J^S$ tensor (denoted as $\|J^S\|$), which contain a combined effect of the two types of couplings, discussed above. In the above equation the site indices are omitted.

\begin{figure}[!h]
\includegraphics[width=\columnwidth]{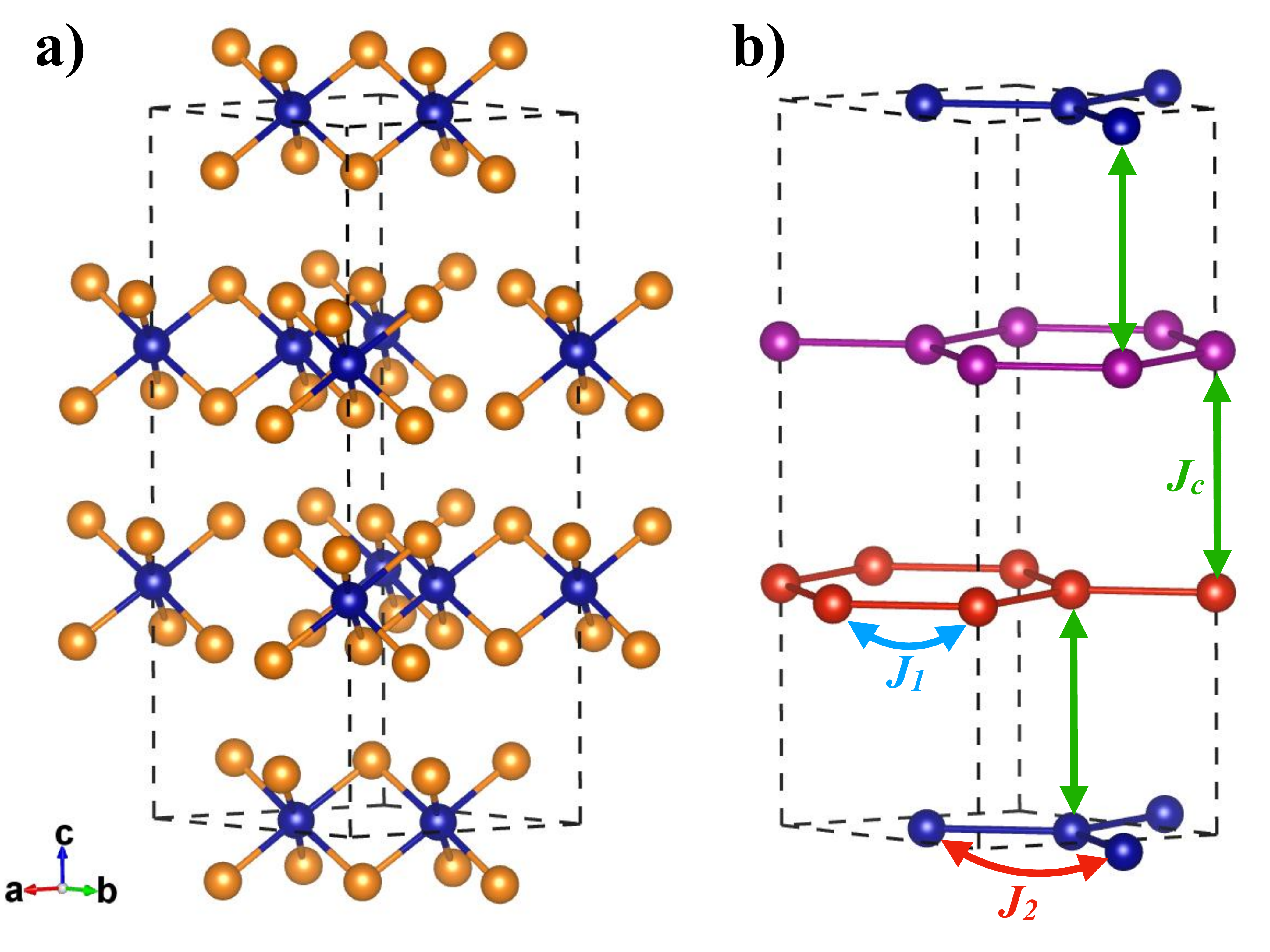} 
\caption{Panel "a": experimental structure for the bulk CrI$_3$ taken from Ref.~\cite{cri3-struct}. Cr(I) atoms are shown as blue (orange) spheres. Panel "b": The networks of Cr atoms along with the most relevant exchange interactions. The Cr atoms belonging to different layers are depicted with various colour for better visualization of the stacking.}
\label{cri3-bulk-str}
\end{figure}

The low-temperature rhombohedral (R$\bar{3}$) structure of CrI$_3$ is shown in Fig.~\ref{cri3-bulk-str}
Each plane consists of honeycomb lattices of Cr atoms, surrounded by six iodine atoms, forming edge-sharing octahedra. 
The Cr layers are ABC-stacked in such a way that every atom in one layer is located right above and below the holes of the honeycomb Cr networks of the adjacent layers.
We defined the structure using hexagonal cell, containing three layers with six Cr atoms in total.

\begin{figure}[!h]
\includegraphics[width=\columnwidth]{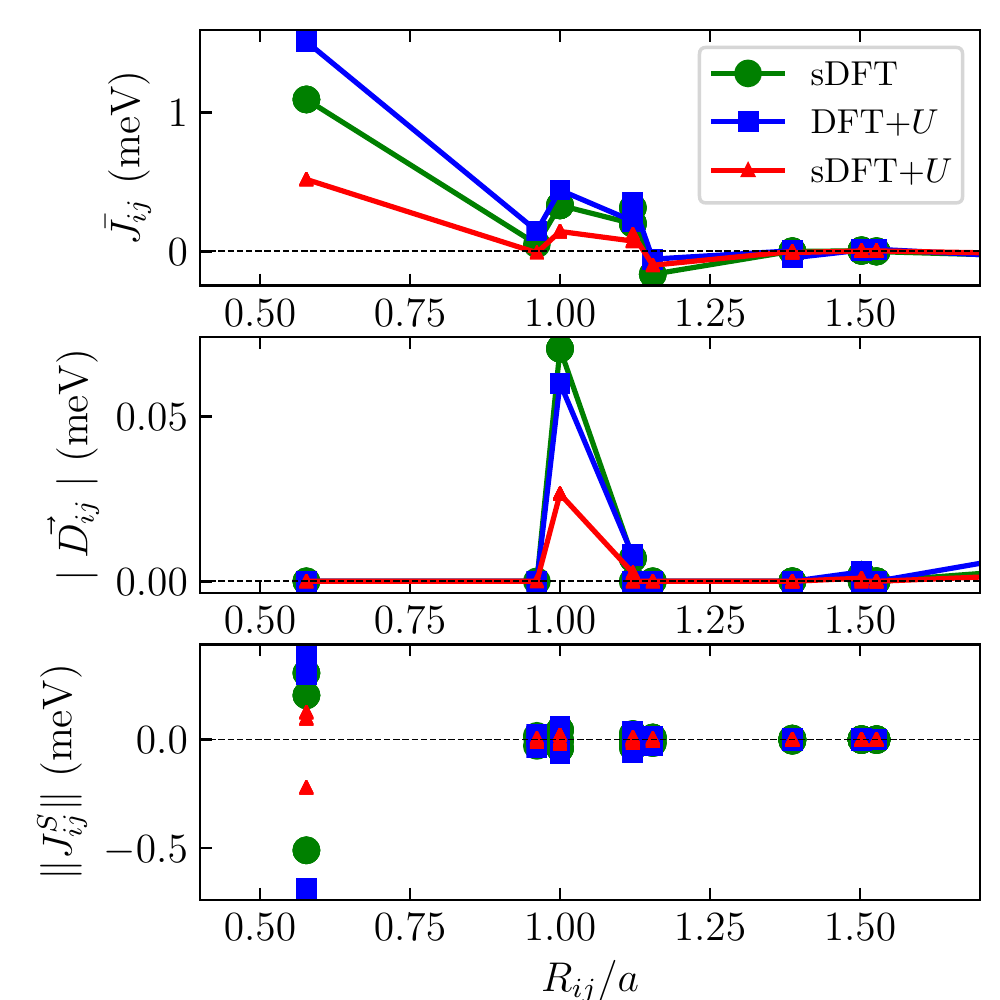} 
\caption{Calculated isotropic and anisotropic exchange parameters as a function of the distance in bulk CrI$_3$.}
\label{jijs-cri3-bulk}
\end{figure}

The calculated exchange parameters defined by Eq.~\eqref{def-int} are shown in Fig.~\ref{jijs-cri3-bulk}.
First of all, one can see that all computational setups suggest that the system has an overall tendency to ferromagnetic order.
The isotropic part of the nearest-neighbour (NN) $\bar J_1$ is the dominant one.
Its value depends on whether the $U$ term is taken into account and whether the underlying functional is spin-polarized or not.

The differences in the $\bar J_1$ value, obtained using three computational schemes, can be tracked down to its individual orbital-by-orbital contributions.
As was pointed out in Refs.~\onlinecite{PhysRevB.99.104432, Kashin_2020}, the $\bar J_1$ coupling consists of two competing terms originating from $e_g$ and $t_{2g}$-like orbitals.
Following Ref.~\onlinecite{PhysRevB.99.104432}, we analyse the behaviour of $\bar{J_1}$, employing analytical expression from superexchange theory~\cite{PhysRev.115.2, Kugel__1982}:
\begin{eqnarray}
\bar{J_1}\approx\biggl(\frac{1}{\Delta^{uu}_{te}} - \frac{1}{\Delta^{ud}_{te}}\biggl) t_{te}^2 - \frac{t_{tt}^2}{\Delta^{ud}_{tt}},
\label{j1orb}
\end{eqnarray}
where $\Delta^{ud}_{te}$, for instance, stands for the exchange splitting between the $t_{2g}$ ($t$) spin-up (u) and $e_g$ ($e$) spin-down (d) states, and $t_{te}$ is the hopping integral between the corresponding orbitals.

\begin{figure}[!h]
\includegraphics[width=\columnwidth]{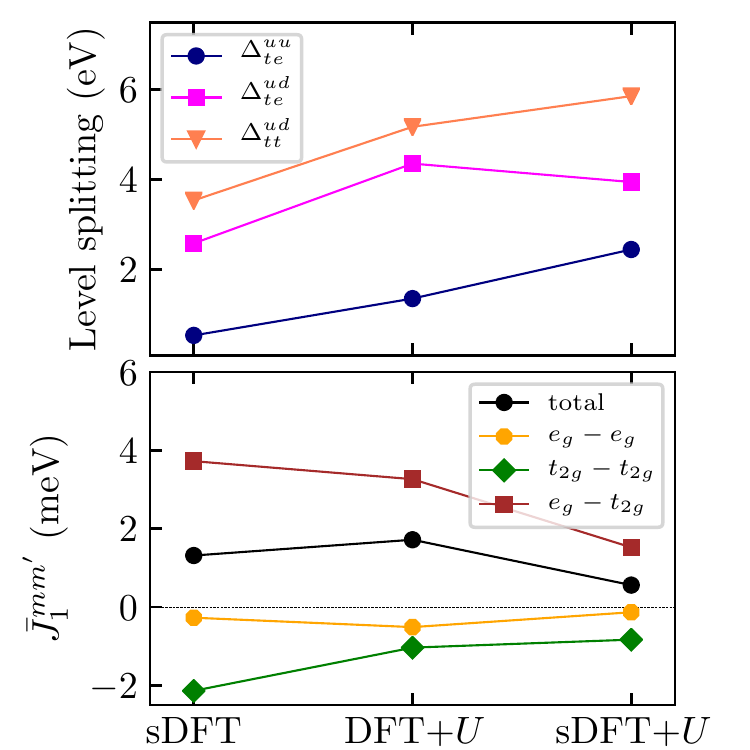} 
\caption{Orbital structure of the NN isotropic exchange interaction in bulk CrI$_3$. Top panel: values of energy splittings between different kinds of $e_g$ and $t_{2g}$ orbitals having spin-(u)p and spin-(d)own projections. Bottom panel: Orbital-resolved NN exchange interaction in bulk CrI$_3$.}
\label{cri3-orb-str}
\end{figure}

Neglecting SOC in the calculation (so that spin becomes a good quantum number), we have extracted the energy splittings between the two effective types of orbitals.
These results are shown in Fig.~\ref{cri3-orb-str} along with the calculated orbital-resolved components of the isotropic part of the NN exchange interaction.
First of all, it is seen that the $\Delta^{ud}_{tt}$ increases as one compares sDFT, DFT+$U$ and sDFT+$U$ extracted values. 
This is directly influencing the $t_{2g}-t_{2g}$ contribution to $\bar{J_1}$, which shows the opposite trend, as expected from Eq.~\eqref{j1orb}.
The situation is more complex with the $e_g-t_{2g}$ contribution. 
Here we find that inclusion of $U$ leads to the increase of both $\Delta^{uu}_{te}$ and $\Delta^{ud}_{te}$ parameters.
However, comparing DFT+$U$ and sDFT+$U$, the two relevant energy splittings behave in a different way.
The crystal-field splitting ($\Delta^{uu}_{te}$) tends to increase much more in sDFT+$U$ as compared to DFT+$U$, whereas the exchange splitting ($\Delta^{ud}_{te}$), on the contrary, goes slightly down.
Both of these factors contribute to an overall decrease of the FM $e_g-t_{2g}$ contribution.
This fact once again highlights the importance of the comparative studies between DFT+$U$ and sDFT+$U$ calculations.

The NN symmetric and next NN antisymmetric anisotropic couplings are not negligible, but still smaller than the isotropic $\bar J_1$.
The NN DM vectors are zero by symmetry and the next NN ones are already not that big.
On top of that, it is not entirely oriented along $z$ axis, which is preferable for opening the topological gap in the spin wave spectrum.
Our results show that $D^z$ is equal to 0.05, 0.02 and 0.012 meV in sDFT, DFT+$U$ and sDFT+$U$, respectively. 
So the $U$ term also acts on the orientation of the DM vectors in a non-trivial way.
Overall, comparing different setups, the results for the $\mid D\mid$ follow the same trend as for $\bar{J}$.

\begin{figure}[!h]
\includegraphics[width=0.9\columnwidth]{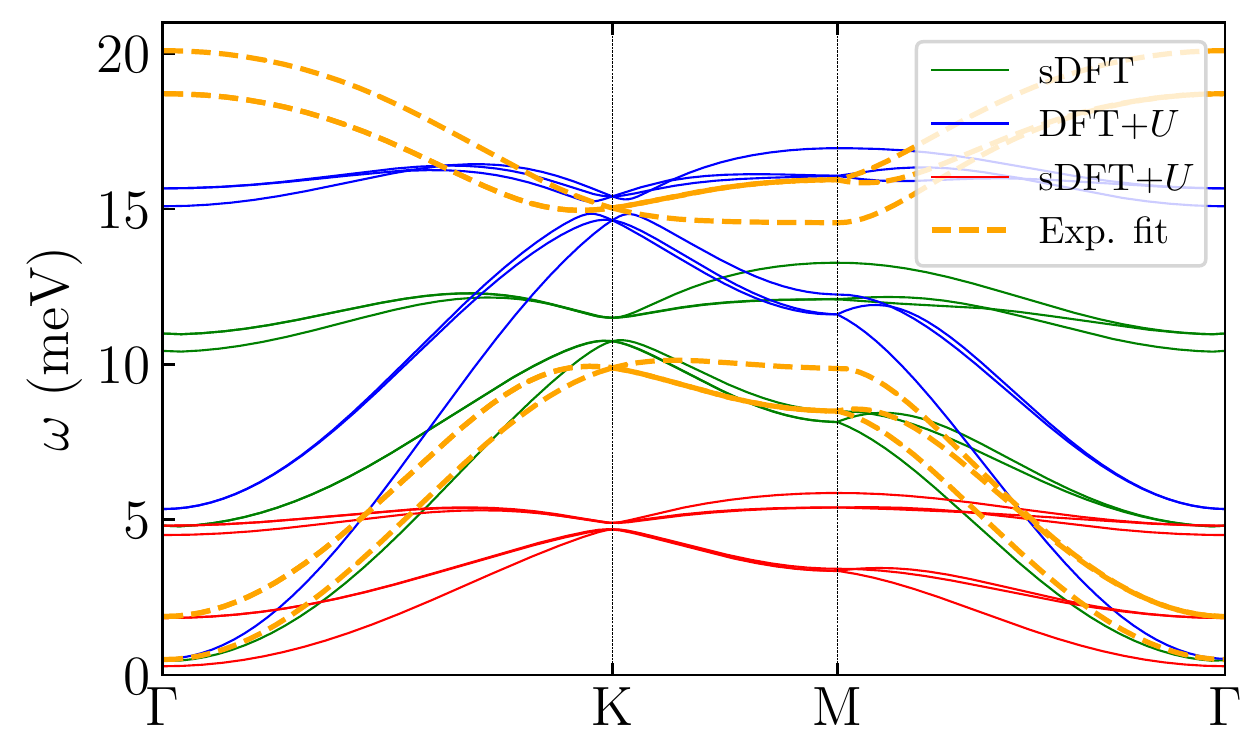} 
\caption{Simulated adiabatic magnon spectra for bulk CrI$_3$ using the parameters shown in Fig.~\ref{jijs-cri3-bulk}.}
\label{cri3-bulk-magnons}
\end{figure}

Using the obtained interaction parameters, we have calculated the adiabatic magnon spectra, shown in Fig.~\ref{cri3-bulk-magnons}. 
Note that we observe six magnon branches, which is equal to the number of magnetic atoms in the hexagonal cell, as a result of finite inter-layer magnetic couplings (see Ref.\onlinecite{PhysRevX.8.041028}).
The MAE per Cr atom was estimated to be 0.47, 0.52 and 0.28 meV in sDFT, DFT+$U$ and sDFT+$U$, respectively. 
The gap in the spin wave spectra at the zone center (which is equal to MAE) is primarily defined by the single ion anisotropy constant $K_{zz}$, whereas the contribution of the two-site anisotropies plays a secondary role.
Only in the case of DFT+$U$ these two contributions are of comparable size.
This conclusion is in contradiction with the total-energy difference analysis presented in Ref.~\onlinecite{Lado_2017} where the anisotropic inter-site interactions are suggested to be relatively stronger than the local anisotropy.

In Fig.~\ref{cri3-bulk-magnons} we also show the simulated magnon spectrum obtained using the fitted values from experiment~\cite{PhysRevX.8.041028}.
Comparing the results against the fitted spectrum, one can see that the sDFT and DFT+$U$-derived values produce the magnon spectrum, which lies in a correct energy range. 
The sDFT+$U$ results result in substantially underestimated magnon energies.
The strongest discrepancy is found for the position of the optical branches, which in our calculation happen to lie at lower energies as well as the dispersion at the $K$ point.
In the experiment there is a gap of about 4 meV at this wavevector.
Our calculations take into account both symmetric and antisymmetric anisotropic interactions, which were discussed as possible sources for the gap opening. 
We indeed observe the gap opening, but its value is not bigger than 0.7 meV (as obtained using sDFT). 
By setting to zero either $J^S$ or the DM terms, we identified that the latter one primarily contributes to this effect.
Still, the magnitude of the gap is not as strong as suggested by experimental data. 
There can be several reasons for such behaviour, such as incompleteness of our spin Hamiltonian (Eq.~\eqref{heis}). 
For instance, it might be necessary to augment it with higher-order interactions or consider lattice effects, which we will discuss in Section IV.

\subsection{2D periodic monolayers}
Next we investigated monolayers of CrX$_3$.
The optimized parameters of the crystal structure are shown in Table~\ref{tab:table2}.
The experimental lattice constants for the bulk samples of CrCl$_3$, CrBr$_3$ and CrI$_3$ are 5.96\AA, 6.3\AA and 6.86\AA, respectively\cite{crcl3-struct,crbr3-struct,cri3-struct}.
The results were obtained with PBESol functional and can also be compared with prior results\cite{PhysRevB.98.144411} obtained with PBE. 
One can see that the lattice constants obtained here are closer to the experimental bulk values, which should be an indication that PBESol performs better for modelling of the structural properties of chromium trihalides.

\begin{table*}[!ht]
\caption{\label{tab:table1} The structural parameters for the relaxed CrX$_3$ monolayers obtained in various setups. Here $z_X$ stands for the distance of the halide atom X(=$\{$Cl,Br,I$\}$) away from its nearest Cr plane (located at $z$=0).}
    \begin{tabular}{c|ccc|ccc|ccc}
          &  & CrCl$_3$ & & & CrBr$_3$ & & & CrI$_3$ & \\
            \hline
         &   sDFT &  DFT+U & sDFT+U & sDFT &  DFT+U & sDFT+U & sDFT &  DFT+U & sDFT+U \\
            \hline
$a$ (\AA) &  5.932 & 5.952 & 5.967 & 6.291 & 6.310 & 6.326 & 6.817 & 6.842 & 6.856  \\
$z_X$ (\AA) & 1.320 & 1.329 & 1.332 & 1.419 & 1.429 & 1.431 & 1.541 & 1.554 & 1.557 \\
$d_{\text{Cr-X}}$ (\AA) & 2.324 & 2.336 & 2.341 & 2.479 & 2.492 & 2.497 & 2.690 & 2.705 & 2.710  \\
    \end{tabular}
\end{table*}

The results obtained with various computational setups turned out to be quite similar. 
There is a clear trend that adding the $U$ term slightly increases the equilibrium lattice constant.
One should keep in mind that since the bond angle is close to 90$^{\circ}$ and there are competing FM and AFM contributions to the NN coupling, the magnetic interactions are highly sensitive to any small changes in the position of the halide atom. 
Hence, the fact that sDFT, DFT+$U$ and sDFT+$U$ produced slightly different structures will also contribute to the differences in the calculated interaction parameters.

\begin{table}[!h]
\caption{\label{tab:table2} Calculated magnetocrystalline anisotropy energy (MAE), obtained from the total energy differences.}
    \begin{tabular}{ccc}
    system & setup & MAE (meV)\\
            \hline 
               & sDFT & 0.01 \\
CrCl$_3$ & DFT+$U$ & 0.03 \\ 
               & sDFT+$U$ & 0.02 \\ 
\hline
               & sDFT & 0.16  \\
CrBr$_3$ & DFT+$U$ & 0.15 \\
               & DFT+$U$ & 0.09 \\
\hline
               & sDFT & 0.70 \\
CrI$_3$   & DFT+$U$ & 0.64 \\
               & sDFT+$U$ & 0.32 \\ 
            \hline
    \end{tabular}
\end{table}

\begin{figure*}[!t]
\includegraphics[width=1.9\columnwidth]{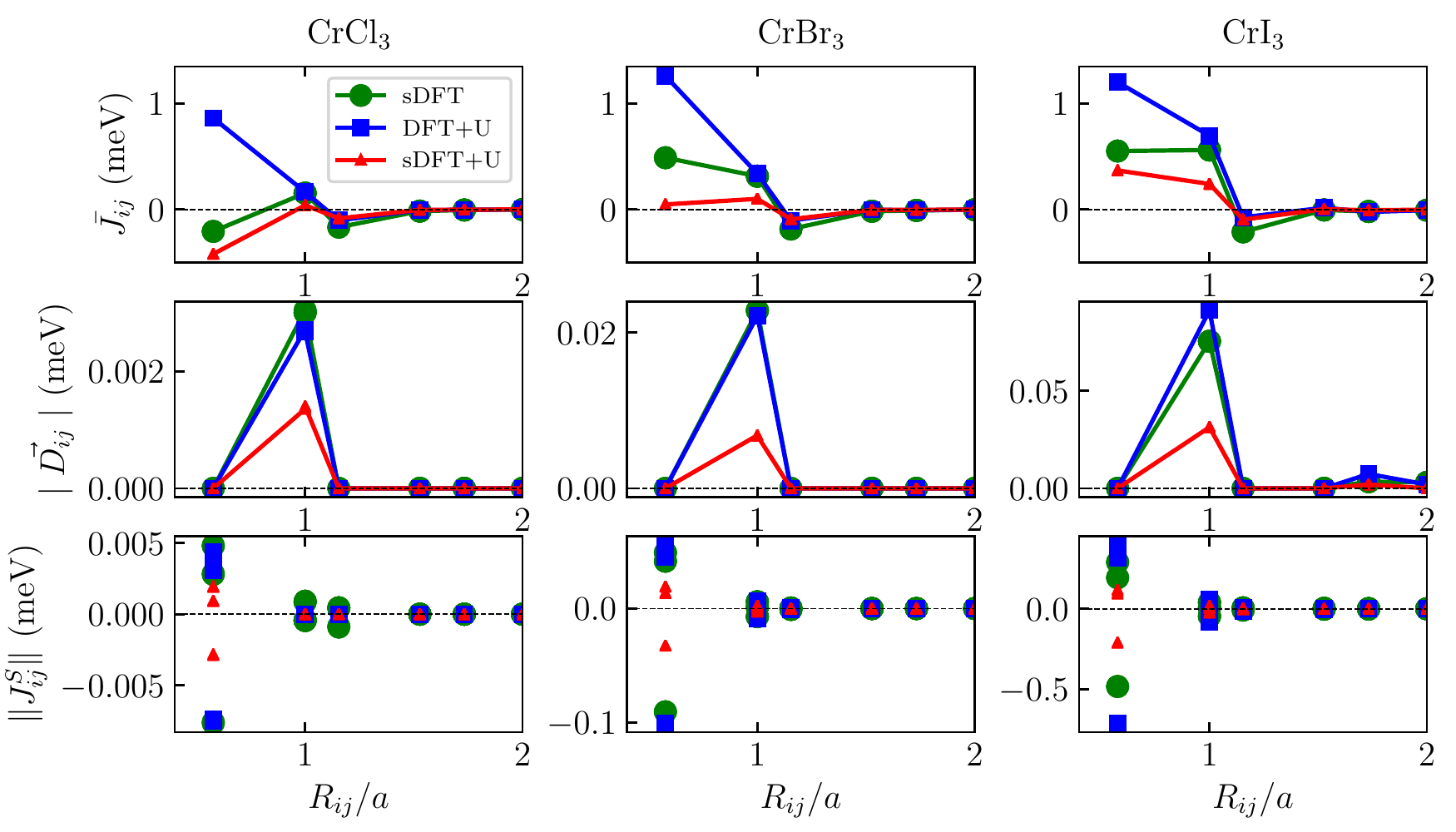} 
\caption{Calculated relativistic exchange parameters as a function of the distance in bulk CrI$_3$. The plotted quantities are defined according to the Eq.~\eqref{def-int}.}
\label{jijs-crx3-mono}
\end{figure*}

The calculated exchange parameters for the series of CrX$_3$ monolayers are shown in Fig.~\ref{jijs-crx3-mono}.
First of all, one can see that the anisotropic interactions ($\vec D$ and $J^S$) scale with the mass of the halide.
This corroborates with the idea that the SOC on the halide atoms plays a crucial role in defining the strength of anisotropic Cr-Cr interactions \cite{Lado_2017}.
The strongest relativistic interactions are found for CrI$_3$. 
For this system, we can also compare the results with respect to those obtained for bulk (Fig.~\ref{jijs-cri3-bulk}).
Overall, the results are quite similar.
The $\bar{J_1}$ becomes smaller, while $\mid \vec D_2 \mid$ gets larger in the single-layer limit as compared to bulk. 
However, this is partially related to the structural relaxations, which were considered for monolayers.

The obtained isotropic exchange couplings in CrI$_3$ can be compared with prior works, where it was also calculated using MFT.
For bulk, we found a very good agreement with prior values\cite{PhysRevB.99.104432}.
Regarding single-layer CrI$_3$, the agreement with Ref.~\onlinecite{Kashin_2020} is less good, but the structural optimization was done using different functionals in these two works, which makes it difficult to compare.
Nonetheless, there is one substantial qualitative difference related to the effect of Hubbard $U$. 
The results of Kashin \textit{et al.} suggest that $J_1$ value is larger in sDFT+$U$ as compared to sDFT.
Here we find the same trend, as was already reported for the bulk CrI$_3$, namely that sDFT+$U$ results in smaller values of the exchange.
The present behaviour is in agreement with the analysis from Ref.~\onlinecite{PhysRevB.99.104432}.

The results for ${\bar J}$'s CrCl$_3$ also suggest an instability of the ferromagnetic state with the value of $\bar J_1$ being negative in the sDFT and sDFT+$U$ schemes. The latter result is in agreement with Ref.~\onlinecite{PhysRevB.99.104432}, where the corresponding bulk of this material was investigated.
Thus only the results obtained with DFT+$U$ are able to provide the correct sign of this interaction. 
In fact, this is the only scheme which neglects an intrinsic magnetism of the ligand states. 
In some way, it supports the findings of Ref.~\onlinecite{PhysRevB.99.104432}, which suggest that if there is an intrinsic magnetism of halide atoms, then it should also be added to a model describing magnetic interactions.
Our results suggest that by excluding this effect from the calculations in the first place provides a better ground for construction of the minimal Heisenberg model, involving only spins of transition metal atoms. 
Same conclusions were reached for transition metal oxides in Ref.~\onlinecite{PhysRevB.97.184404}.

Next, we address the MAE, which values are shown in Table~\ref{tab:table2}. 
The results are in good agreement with Refs.~\onlinecite{PhysRevB.98.144411, C5TC02840J}, even though another functional have been used in those two works.
One can see that the heavier the halide, the stronger the tendency to the out-of-plane orientation of the moments.
This fact has also been observed experimentally for the series of Cr(Cl$_x$,Br$_{1-x}$)$_3$ compounds\cite{abramchuk2018}.
Prior DFT calculations, based on VASP, predicted the monolayer of CrCI$_3$ to have weak out-of-plane anisotropy\cite{PhysRevB.98.144411, C5TC02840J}.
We indeed reproduce this result by employing full-potential electronic structure method in this work.

\begin{figure}[!h]
\includegraphics[width=0.9\columnwidth]{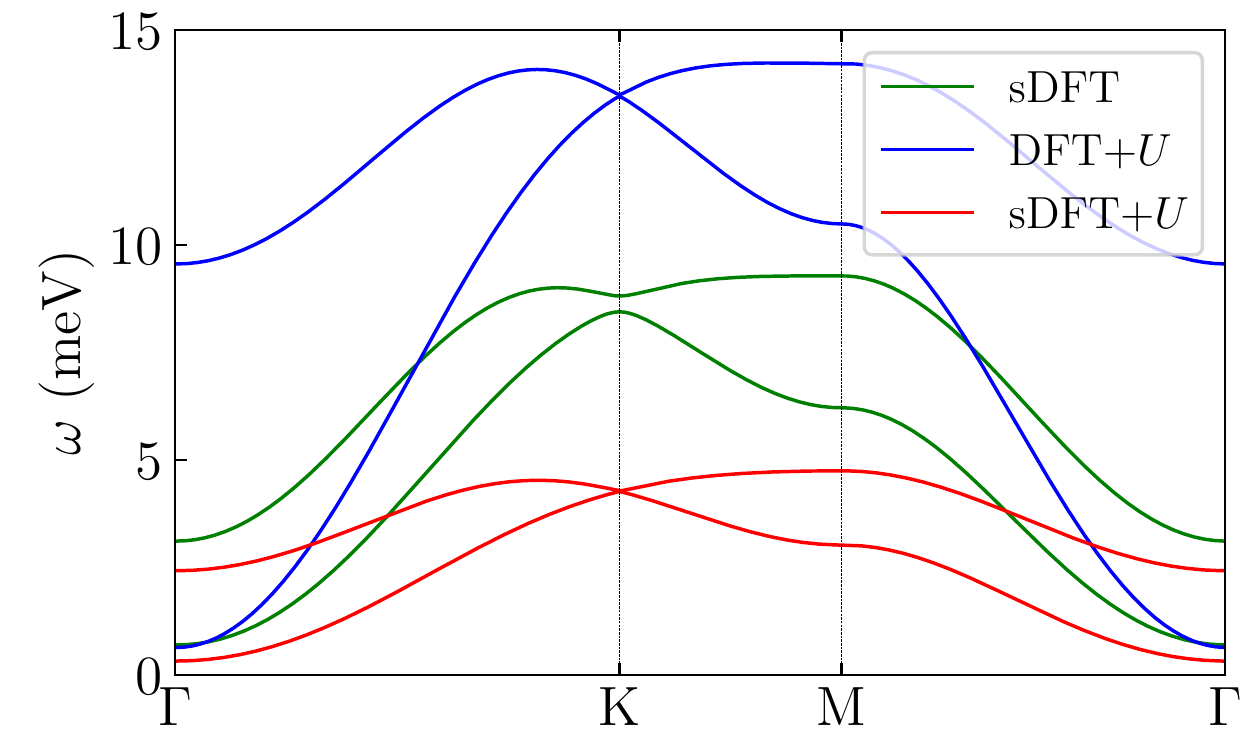} 
\caption{Simulated adiabatic magnon spectra for monolayered CrI$_3$ using the parameters, obtained for three computational setups, shown in Fig.~\ref{jijs-crx3-mono}.}
\label{cri3-mono-magnons}
\end{figure}

\begin{figure}[!h]
\includegraphics[width=0.9\columnwidth]{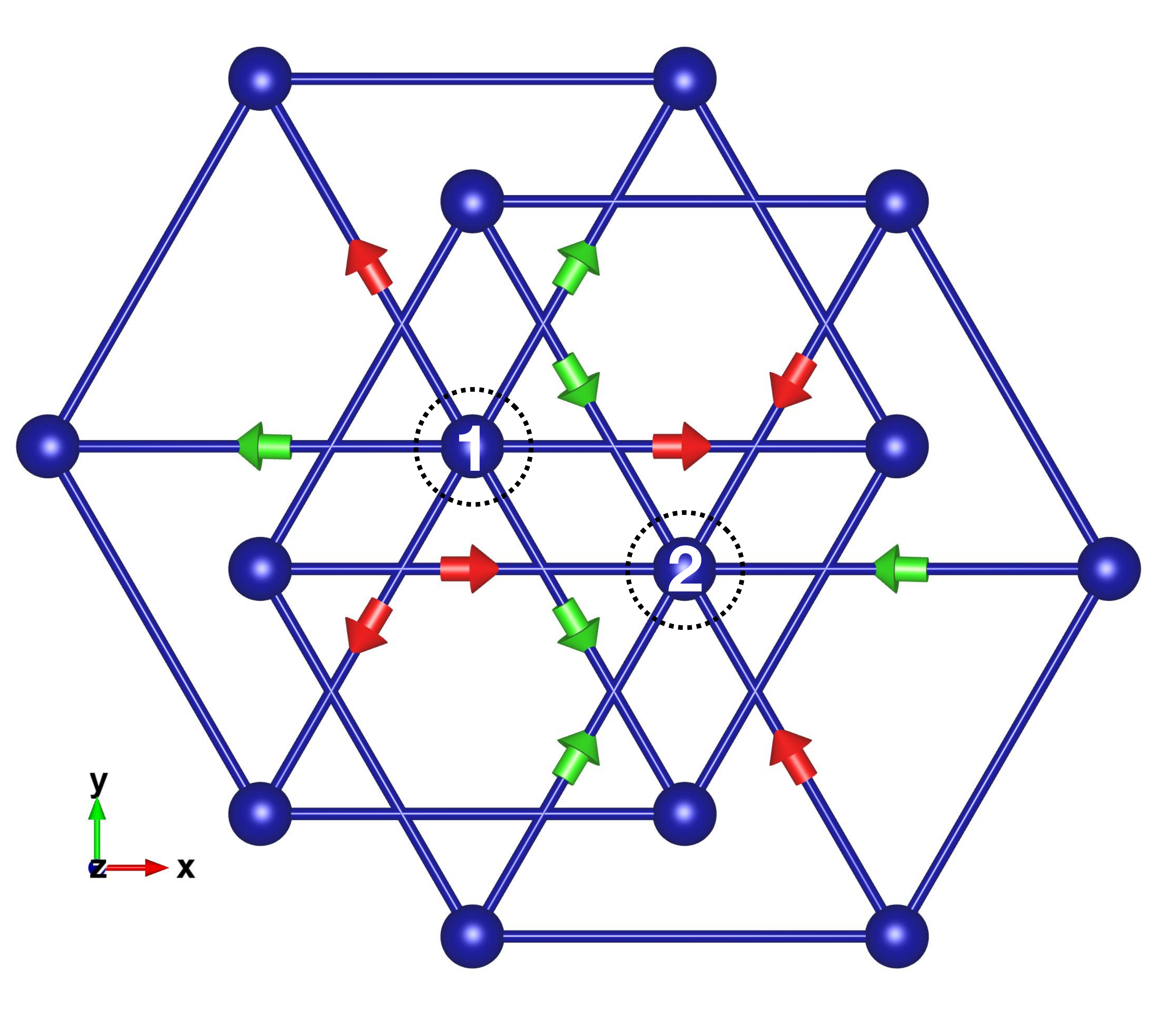} 
\caption{Calculated next NN $\vec D_{1j}$ and $\vec D_{2j}$ vectors for the two Cr sublattices (1 and 2) in monolayered CrI$_3$. DFT+$U$ derived data is presented. The colour corresponds to the sign of a small $z$-component of the DM vector.}
\label{dmi-mono}
\end{figure}

Using the obtained interaction parameters, we have calculated the adiabatic magnon spectra, which are presented in Fig.~\ref{cri3-mono-magnons}. Comparing with bulk (Fig.~\ref{cri3-bulk-magnons}), we find that overall the spin waves are predicted to get softer in the monolayer limit.
The strongest difference is in the value of the topological gap, which becomes negigible for the (s)DFT+$U$ case.

Why is the gap at the K point much smaller than the value of next NN DM vector, even though they should be related ?
Fig.~\ref{dmi-mono} depicts the directions of the obtained DM vectors in monolayered CrI$_3$. 
One can see that they are oriented nearly along the corresponding Cr-Cr bonds and obey the global symmetry rules.
The sublattices are clearly different, since all $\vec D_{1j}$ vectors for the first sublattice are pointing away from the atom, whereas it is the opposite for the atom belonging to the other sublattice.
The DM vectors are lying primarily in the plane of the Cr networks, whereas it is the $D^z$ component responsible for the opening of the gap at the K point.
In sDFT and DFT+$U$ calculations, the $D^z$ was estimated to be 0.02 and 0.01 meV, respectively.
As a result, the DM interactions in monolayers only marginally affect magnon band splitting at the K point.

To conclude, according to our results, the anisotropic interactions indeed lead to the gapped topological magnon states in CrI$_3$.
However, the gap between the two magnon branches is even smaller than the value we obtained for the bulk of this material.

\subsection{Finite-size ribbons}
Even though the gap at the point $K$ of the BZ is small compared to experiment, the qualitative picture of two topological modes is reproduced by our calculations.
Such topology of the magnons implies that there will be surface/edge modes, having chiral nature and being robust against perturbations, which makes them very promising for magnonic technology. 
However, the dynamics of the magnon edge modes can be strongly affected by any modification of the interaction parameters, which can be expected to appear at the edge of the material.

\begin{figure}[!t]
\includegraphics[width=0.9\columnwidth]{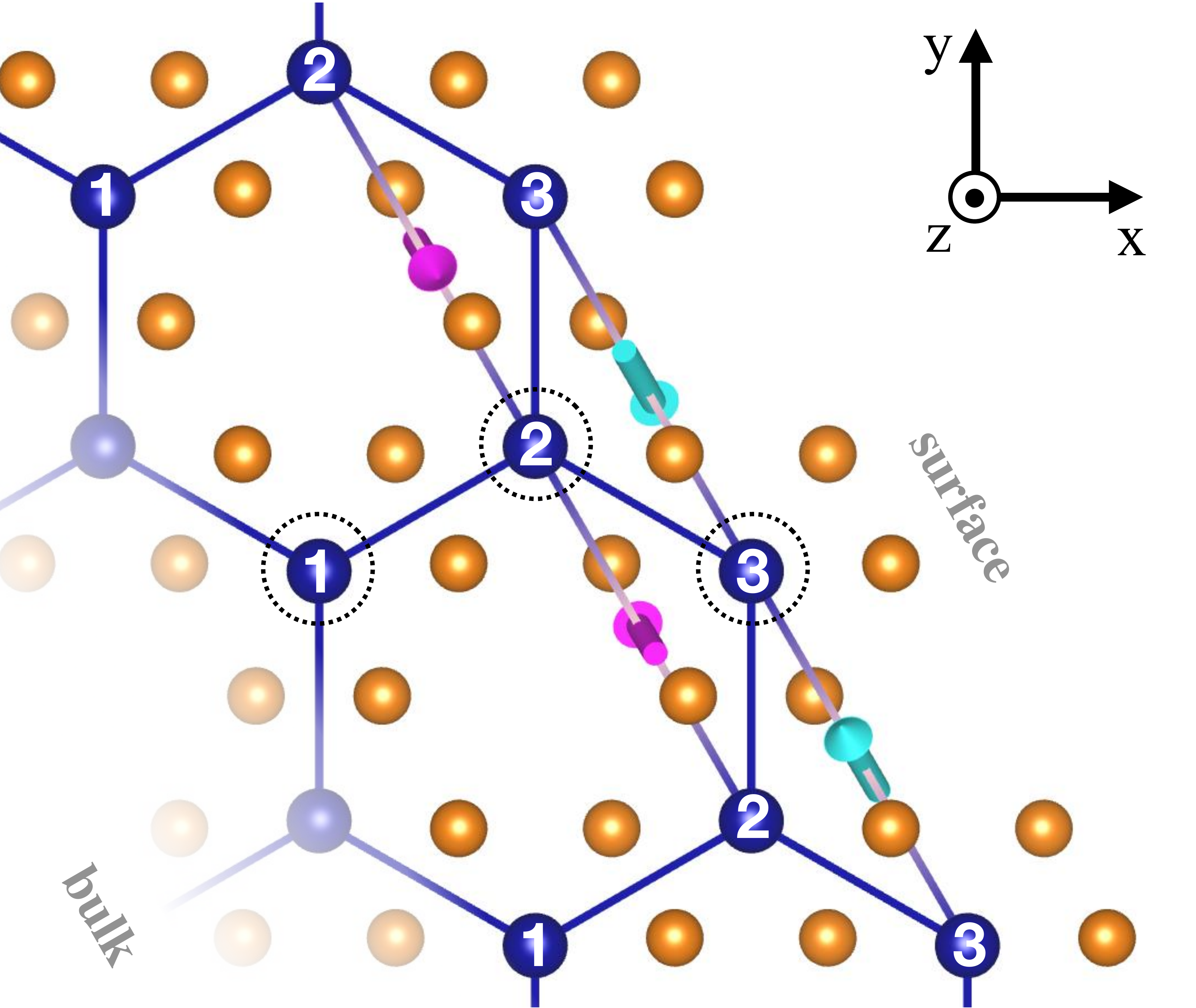} 
\caption{Schematic picture of the simulated finite-size flake of CrI$_3$. The Cr atoms closest to the edge are labeled. The directions of the largest DM vectors are shown with arrows. The dashed circles highlight the atom the DM vectors are calculated for.}
\label{dmi-ribbon}
\end{figure}

For this purpose we have broke the periodicity along one of the directions in the 2D plane and added a 20$\AA$ thick vacuum.
We have preserved octahedral environment for all Cr atoms including the ones at the edge of the 1D ribbon. 
We believe that this structure should resemble experimental situation, where the Cr dangling bonds will be saturated in order to screen its charge. 
The resulting structure is shown in Fig.~\ref{dmi-ribbon}.
For our analysis, we have selected three types of chromium atoms located close to the surface of the ribbon.
Fig.~\ref{dmi-ribbon} shows the magnetic interactions between these atoms.

\begin{figure}[!h]
\includegraphics[width=0.9\columnwidth]{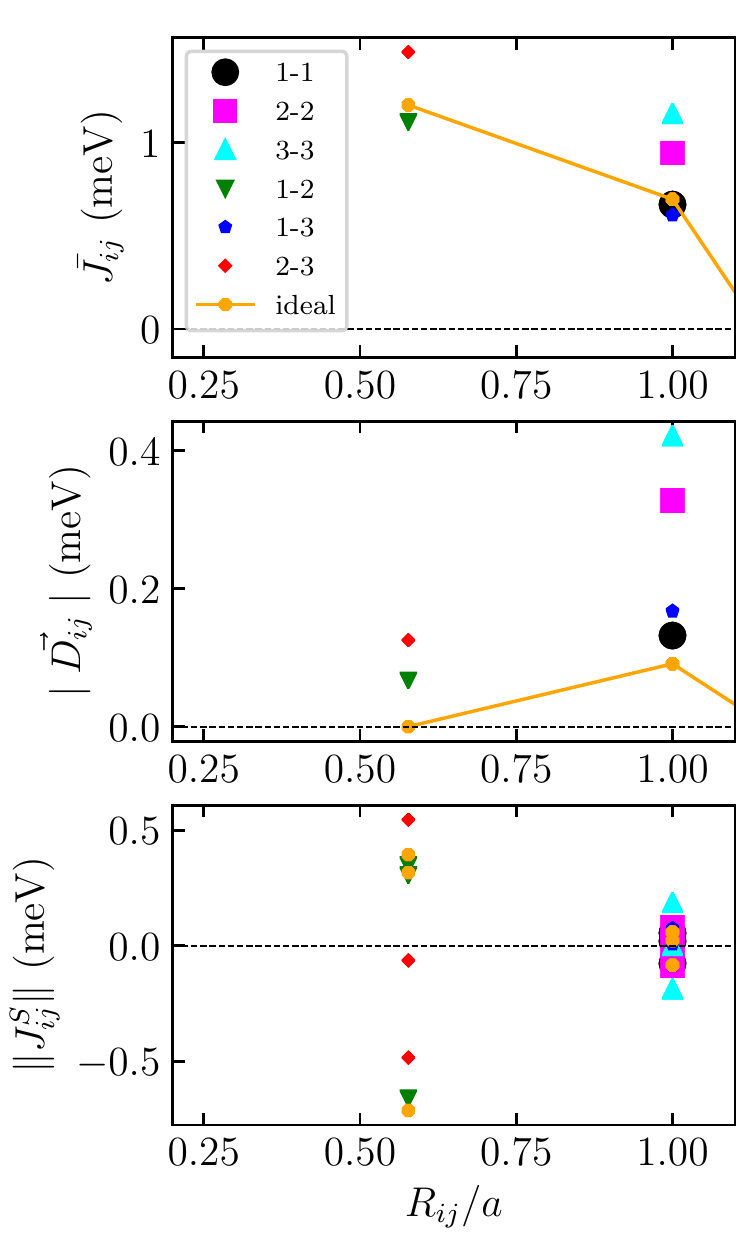} 
\caption{Exchange interactions between selected Cr atoms located the closest to the edge of the material. The results for DFT+$U$ are presented along with the corresponding values obtained for ideal periodic 2D monolayer.}
\label{jijs-ribbon}
\end{figure}

The values of the Cr magnetic moments are close to the nominal 3 $\mu_B$ and are practically the same for all types of Cr atoms.
Since the inversion symmetry between the nearest neighbours is broken by the presence of the vacuum, the corresponding DM vector becomes finite for the atoms close to the edge. 
Still, the $\vec D_2$ is the dominant interaction among DM vectors.
Moreover, there are also less symmetry constraints for the DM vectors for the atoms at the edge of the ribbon. 
As a result, these interactions between Cr(3) atoms become nearly four times stronger than those in the bulk. 
Their orientation also changes and they start pointing more away from the surface, which should influence the dynamics of the edge spin waves.
Analysing the interactions between the atoms further from the edge, one can see that the finite-size effects are quite short-ranged.
Cr(1) atoms already experiences exchange interactions similar to those in the periodic monolayer, meaning that the electronic structure is not influenced by the presence of the edge.
The results for $J^S$ suggest that the symmetric anisotropic interactions and Kitaev exchange are much less sensitive to the proximity to the edge as compared to the DM vectors.

\section{Outlook}

We have addressed the full tensor of exchange interactions between Cr ions in bulk CrI$_3$ as well as in the series of monolayered chromium trihalides.
The calculations for bulk CrI$_3$ correctly predict the gap opening at the $K$ point.
However, the magnitude of the gap was found to be much smaller than in experiment.
Both DM interactions and Kitaev-type were found not to be sufficiently strong to reproduce the gap of $\approx$4 meV, seen in the experiment\cite{PhysRevX.8.041028}.
The next NN DM vectors were also found to lie primarily in plane, which does not contribute to the corresponding mode splitting.

The magnetic interaction tensors $J^{\alpha\beta}_{ij}$ for the bulk and monolayer CrI$_3$ were also compared.
The isotropic coupling was found to decrease in the single layer limit, which is consistent with the experimentally observed decrease of $T_c$ as compared to bulk\cite{Huang2017}.
Analysing the results for the series of CrX$_3$ compounds we find that the SOC-related properties, such as MAE, DM interaction and symmetric anisotropic exchange all scale with the mass of halide atoms.
It is shown that the effect of Hubbard $U$ on the dominant ferromagnetic coupling depends on the underlying spin-polarization of the xc functional. 
This originates from the fact that the resulting crystal fields and the exchange splittings between various orbitals are different in DFT+$U$ and sDFT+$U$ methods.
All three considered monolayers of chromium trihalides are predicted to be ferromagnetic only within DFT+$U$ setup, where the intrinsic magnetism of ligand orbitals is neglected.

Finally, we investigate the properties of ribbon of CrI$_3$ trying to quantify the impact of finite-size effects on the magnetic interactions.
The NN DM couplings, which are forbidden in the periodic structure, become finite and gain substantial values close to the edges of the material.
In addition, there is a strong (factor five) enhancement of the next NN DM interactions.
We suspect that the dynamics of the edge magnon states might be strongly affected by the produced changes in the $J^{\alpha\beta}_{ij}$'s, which should be done in a separate study.

The disagreement between the present theory and experiment for bulk CrI$_3$ might originate from the lattice effects.
The latter is known to be able to open the gaps in the spin wave spectra\cite{Turov_1983}.
Note that phonon-induced intervalley scattering can lead to a dynamical 
gap opening at K and K' points in the electron spectrum of graphene\cite{PhysRevB.75.155420,KANDEMIR201480}. 
One can assume that a similar effect can take place in magnetic honeycomb lattice as well, but this question needs further investigation.

Strong spin-phonon coupling has been reported for CrX$_3$ systems \cite{C8CP03599G, PhysRevMaterials.1.014001}.
Since phonons and magnons in this class of materials have similar energies~\cite{C8CP03599G, C5TC02840J}, even a small coupling can produce a substantial hybridisation between the two types of excitations.
One can expect the exchange interactions will be sensitive to any structural distortions, due to the NN bond angle being close to 90$^{\circ}$ and the internal competition between FM and AFM contributions.
This is indirectly supported by the fact that the strain ($\epsilon$) in monolayer CrX$_3$ was shown to lead to substantial changes of the effective NN exchange\cite{PhysRevB.98.144411}. 
In particular, for CrCl$_3$ it was predicted even that this coupling can change its sign for $\epsilon$>2.5$\%$.

The coupled dynamics of phonons and magnons can currently be parameterized from first-principles calculations and then solved exactly in the limit of classical spins\cite{PhysRevLett.121.125902, PhysRevB.99.104302} or approximately in quantum flavor\cite{PhysRevB.100.014430}.
We foresee that such investigation for CrX$_3$ systems will be done in the nearest future.

Another possible reason is the neglect of higher-order (e.g. biquadratic or ring) exchange interactions, which are sometimes pronounced in $3d$-metal-based systems\cite{pnictides-biq,antropov-nat11,fedorova-biquad}. We do not think however that this is the point since the magnetic interactions found from small deviations via MFT take into account all these effects, and the magnon excitation spectra are directly related to these magnetic parameters\cite{KatsLich2004}.
Finally, there is also a possibility that the magnetic Hamiltonian involving only Cr spins is not sufficient to describe all the properties of these materials and that an explicit account of the halide moments is essential. 
The latter idea was proposed by Besbes \textit{et al.}\cite{PhysRevB.99.104432}.

\begin{acknowledgments}
YOK acknowledges the financial support from the Swedish Research Council (VR) under the project No. 2019-03569. The work of MIK is supported by JTC-FLAGERA Project GRANSPORT. 
\end{acknowledgments}


%

\end{document}